\documentclass[12pt]{iopart}

\usepackage{graphicx, subfigure}
\usepackage{amstext, iopams}
\usepackage{hyperref}

\begin{document}

\title[Phase-field model for GB grooving]{Phase-field model for grain boundary grooving in multi-component thin films}

\author{Mathieu Bouville$^1$, Shenyang Hu$^2$, Long-Qing Chen$^3$, Dongzhi Chi$^1$ and David J Srolovitz$^{1,4,5}$}

\address{$^1$ Institute of Materials Research and Engineering, Singapore 117602}
\address{$^2$ MST-8, Los Alamos National Laboratory, Los Alamos, NM 87545, U.S.A.}
\address{$^3$ Department of Materials Science and Engineering, Pennsylvania State University, University Park, PA 16802, U.S.A.}
\address{$^4$ Institute for High Performance Computing, Singapore 117528}
\address{$^5$ Department of Mechanical and Aerospace Engineering, Princeton University, Princeton, NJ 08544, U.S.A.}
\ead{m-bouville@imre.a-star.edu.sg}

\begin{abstract}
Polycrystalline thin films can be unstable with respect to island formation (agglomeration) through grooving where grain boundaries intersect the free surface and/or thin film-substrate interface. We develop a phase-field model to study the evolution of the phases, composition, microstructure and morphology of such thin films. The phase-field model is quite general, describing compounds and solid solution alloys with sufficient freedom to choose solubilities, grain boundary and interface energies, and heats of segregation to all interfaces. We present analytical results which describe the interface profiles, with and without segregation, and confirm them using numerical simulations. We demonstrate that the present model accurately reproduces the theoretical grain boundary groove angles both at and far from equilibrium. As an example, we apply the phase-field model to the special case of a Ni(Pt)Si (Ni/Pt silicide) thin film on an initially flat silicon substrate.  
\end{abstract}

\vspace{2pc}

\submitto{\MSMSE}

\section{\label{sect_introduction}Introduction}
While grain boundary grooving as a source for agglomeration in thin films is well known \cite{Srol-JOM}, most theoretical models for grooving-induced agglomeration are too idealized to be useful to predict agglomeration in technologically interesting materials.  Most such materials are alloys containing many elements and often multiple phases.  In such materials, segregation of one or more of the alloy components to one or more of the interfaces (surface, grain boundaries, film/substrate interface) is the rule, rather than the exception.  Hence, models based upon grain boundary grooving in single component films with no intermixing between the film and substrate are useful to understand the phenomenon of grain boundary grooving-induced agglomeration but cannot be relied upon to predict the rate of grooving or the phenomenology that has its origin in alloying effects.  Unfortunately, including the full richness of such phenomena in real materials necessarily forces us to employ numerical, rather than analytical, treatments.  In this paper, we describe a phase-field-based method for modelling the evolution of grain boundary grooves that is sufficiently general to include interfacial segregation, compound formation, and interdiffusion.

The grain boundary grooving phenomenon has been known and studied theoretically for some time \cite{Bailey-PPSB-50,Mullins,srol-JAP-86,miller-JMR-90,Nolan-JAP-92}. Grooving is a general phenomenon which occurs at the free surface of all polycrystalline materials.  While the model for grain boundary grooving described below is quite general, for specificity we focus on a particular example: a polycrystalline metal silicide film on a silicon substrate.
The formation and evolution of thin polycrystalline metal silicides or germanides 
is key to the performance of many semiconductor devices.
However, these films can be unstable at high temperature through agglomeration into silicide islands {\it via} grain boundary grooving \cite{Lee-microelec-02}. The films may then lose connectivity. Experimental observations suggest that alloying NiSi with Pt \cite{Detavernier-APL-04,Lavoie-microelec-04} or implanting BF$_2$ \cite{Lavoie-microelec-04,Wong-APL-02} delays or suppresses NiSi agglomeration.
Since segregation of platinum or fluorine may play a role in this phenomenon, our model will include the possibility for one or more species to segregate to a grain boundary and/or interface.

The phase-field approach has been extensively used for predicting microstructure evolution in different materials processes; for example, solidification, ferroelectric transformations in thin films, spinodal decomposition, grain growth, and dislocation dynamics \cite{Chen-JOM-96,Chen-ARMR-02,Karma-encyclopedia-01}. Since our focus in the grain boundary grooving/thin film agglomeration modelling is on a scale that is large compared to the interatomic distance, yet small compared with the grain size, the phase-field method provides an excellent means through which to model the microstructure/morphology evolution.

In this paper, we develop a phase-field model to describe the evolution of the microstructure and morphology of a polycrystalline thin film alloy, including atomic diffusion, phase transformation, grain boundary grooving and morphology evolution.  To this end, we construct a general free energy function that allows for an arbitrary number of grains, phases and atomic constituents.  Analytical results for the profiles of the order parameters at a grain boundary are presented within the framework of this free energy model.  We also discuss the numerical procedure used for evolving the phase fields.  Finally, we present a series of results on grain boundary grooving and Pt segregation in a polycrystalline Ni$_{1-x}$Pt$_x$Si film on Si.

\section{\label{sect_method}Description of the phase-field method}
\begin{figure}
	\centering
	\includegraphics{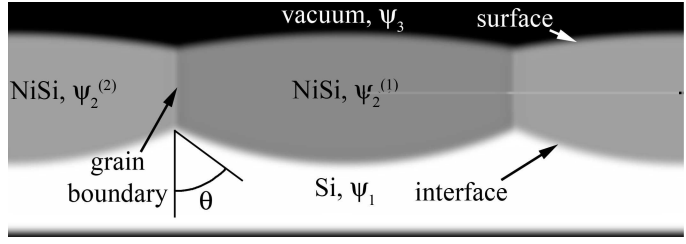}
	\caption{\label{geometry}A schematic illustration of a polycrystalline NiSi film on a silicon substrate. The \{$\psi$\} are the order parameters. The definition of the groove angle $\theta$ is also shown.}
\end{figure}

We consider the case of a polycrystalline film above which is a vacuum and below which is a substrate. Figure~\ref{geometry} is an illustration for the case of a NiSi film on Si.  The model presented below also allows for the formation of other phases (e.g., NiSi$_2$).

To describe grain boundary grooving and atom diffusion during microstructure evolution (we also allow for phase transitions), we use two sets of phase-field variables in our model. One set is order parameter fields which describe phases, while the other set describes composition fields. In order to describe compounds, we distinguish between different sublattices. We define $c_i^{(j)}$ as the concentration of species $j$ on sublattice $i$. In the case of the Ni-Pt-Si-Ge system, sublattice 1 refers to the metals, which may substitute for one another, and sublattice 2 refers to the group-IV elements: on the metal  sublattice $c_1^{(1)}$ and $c_1^{(2)}$ are the concentrations of Ni and Pt and on the group-IV sublattice $c_2^{(1)}$ and $c_2^{(2)}$ are the concentrations of Si and Ge.
The order parameters ${\psi_k}$ indicate the fraction of phase  $k$ present. Each phase $k$ may be polycrystalline, $\psi_k^{(m)}$ represents grain $m$ of phase $k$. In what follows the index $i$ will always refer to types of atoms (i.e., sublattices), $j$ to species, $k$ to phases, and $m$ to grains.

In phase-field theory, the rate of evolution of the system is related to the variation of the total free energy of the system. The evolution of the conserved fields (i.e., compositions) is routinely described through the Cahn-Hillard equation~\cite{Cahn-acta_met-61}
\begin{equation}
	\frac{\partial\, c(\mathbf{r}, t)}{\partial\, t} = \mathbf{\nabla} M\, \mathbf{\nabla} \frac{\delta\, G}{\delta\, c(\mathbf{r}, t)}.
\label{Cahn-Hillard}
\end{equation}
\noindent The non-conserved order parameter fields are governed by the Allen-Cahn equation~\cite{Allen-Jphys-77}
\begin{equation}
	\frac{\partial\, \psi(\mathbf{r}, t)}{\partial\, t} = -L \frac{\delta\, G}{\delta\, \psi(\mathbf{r}, t)}.
\label{Allen-Cahn}
\end{equation}
\noindent In these equations $G$ is the total free energy of the system, $c$ is a concentration (conserved field), $\psi$ is a non-conserved field (phase), and $M$ and $L$ are atom and interface mobilities respectively. $\delta\, G /\delta\, c$ and $\delta\, G /\delta\, \psi$ are functional derivatives of $G$ with respect to $c$ and $\psi$, respectively.

\section{\label{sect_G}Total free energy}
In order to apply (\ref{Cahn-Hillard}) and (\ref{Allen-Cahn}), we first need an expression for the free energy of the system that includes all of the physics necessary to describe the agglomeration process in polycrystalline thin films. Such free energies have been obtained for many  different physical situations \cite{Lai-PRB-90,Braun-roy_socA-97,Braun-acta_mat-97,Wang-acta_mat-98}. The present derivation of the free energy is based upon some of these earlier constructions.

Ignoring elastic contributions to the free energy for now, the  total free energy of the system can be expressed as
\begin{eqnarray}
	G \left(\{\psi\}, \{c\}\right) = \int_V &h \left(\{\psi\}, \{c\}\right) - s \left(\{\psi\}, \{c\}\right) T +\nonumber\\
			&\sum_{k} \kappa_{k} \sum_{m} \left\|\mathbf{\nabla} \psi_k^{(m)}\right\|^2+
\sum_{i} \sum_{j} \lambda_{i}^{(j)} \left\|\mathbf{\nabla} c_i^{(j)}\right\|^2\,\rmd V.
\label{G_equation-init}
\end{eqnarray}
\noindent In this expression, T is the temperature and $h \left(\{\psi\}, \{c\}\right)$ and $s \left(\{\psi\}, \{c\}\right)$ are the enthalpy and  entropy of the system, respectively. The integral is over the entire system. While all grains of the same phase $k$ exhibit the same thermodynamic behaviour, each type of atom is unique.  This explains why $\kappa$ depends only upon the phase $k$, while $\lambda$ depends on both the sublattice $i$ and the atomic species $j$.  

We first focus upon uniform phases (i.e., the gradient terms in (\ref{G_equation-init}) are set to zero).  In each  bulk phase, one of the order parameters is unity, while those describing all other phases are zero.  To ensure this, we include a term in the enthalpy, $h_1$, that has minima at $ \psi_k^{(m)}=0$ and $1$ for each $k$ and $m$.  As is commonly done, we use a double-well potential to describe this contribution to the enthalpy:
\begin{equation}
	h_1 = \sum_{k} D_k \sum_{m} \left(\psi_k^{(m)}\right)^2 \left(1-\psi_k^{(m)}\right)^2.
\label{double-well}
\end{equation}

It is also necessary to ensure that the phase of interest corresponds to a minimum in enthalpy at a particular stoichiometry. In our reference application, we consider four phases, Si, MSi, MSi$_2$ and vacuum, where M represents the metal species (either Ni or Pt or some combination of these). In the MSi$_2$ phase, a third of the sites are occupied by metal atoms: $c_\text{Ni}+c_\text{Pt}=1/3$. Therefore, the enthalpy of this phase must be a minimum at $\sum_j c_i^{(j)}=\Xi_i(k)=1/3$ for $i=\text{metals}$ and $k=\text{MSi}_2$. We describe the enthalpy as a quadratic function for each type of species with a minimum at the desired stoichiometry:
\begin{equation}
	 h_2=\sum_i \left[\sum_k A_i(k) \sum_{m} \psi_k^{(m)}\right] \, \left(\sum_j c_i^{(j)} - C_i\right)^2,
	\label{compo-dep_metal}
\end{equation}
\noindent where $C_i$ is the stoichiometry of the system at any point in space: 
\begin{equation}
	C_i = \sum_k \Xi_i(k) \sum_{m} \psi_k^{(m)}.
\label{def-C}
\end{equation}
\noindent $\sum_{m} \psi_k^{(m)}$ is the amount of phase $k$ and $\Xi_i(k)$ represents the stoichiometry of phase $k$. The second parenthesis in (\ref{compo-dep_metal}) is the difference between the local concentration of atoms of type $i$ and stoichiometry. $A_i(k)$ is the solubility of atoms of type $i$ in phase $k$.  Large $A_i(k)$ implies a small solubility of $i$ atoms on the ``wrong'' sublattices in phase $k$.

\begin{figure}
	\centering
	\includegraphics[width=7cm]{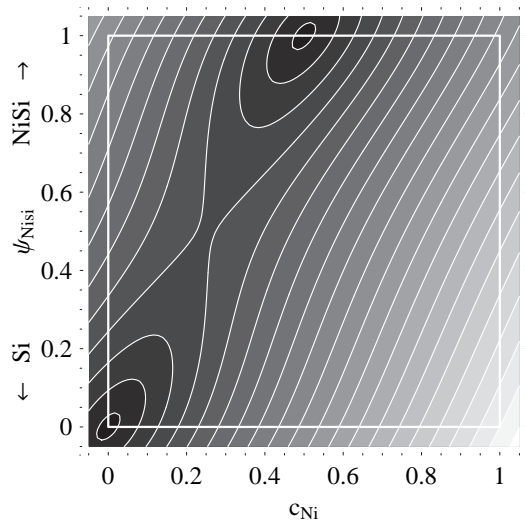}
	\caption{The enthalpy of a Ni-Si system with two phases, (Si) and NiSi, as a function of $c_\text{Ni}$ and $\psi_\text{NiSi}$.}
	\label{energy_map}
\end{figure}

Although the double-well potential of (\ref{double-well}) ensures that each $\psi$ is either 0 or 1 in the bulk phase, it is possible for the phase fractions ($\psi$) of several different phases to be  1 simultaneously (e.g.,  simultaneously 100\% monosilicide and 100\% vacuum).  Clearly, this is unphysical. To rectify this problem, we add a contribution to the enthalpy that contains cross terms between the order parameters:
\begin{equation}
	h_3 =
\sum_{k} \sum_{k^\prime < k} X_{kk^\prime} \left(\sum_m \psi_k^{(m)}\right)^2 \left(\sum_{m^\prime} \psi_{k^\prime}^{(m^\prime)}\right)^2,
\end{equation}
\noindent where the coefficients $X_{k k^\prime}$ are chosen to be sufficiently large so that the total phase fraction at any point is close to unity. A similar expression is used for grain boundaries (i.e., where $k=k^\prime$):
\begin{equation}
	h_4 =
\sum_{k} X_{kk} \left[\sum_m \sum_{m^\prime<m} \left(\psi_k^{(m)}\right)^2 \left(\psi_{k^\prime}^{(m^\prime)}\right)^2\right].
\end{equation}
\noindent Figure \ref{energy_map} shows that the enthalpy of a Ni-Si system with two phases ---(Si) and NiSi--- has a minimum corresponding to the silicon phase at ($c_\text{Ni}=0, \psi_\text{NiSi}=0$) and another corresponding to the NiSi phase at ($c_\text{Ni}=1/2, \psi_\text{NiSi}=1$).

There is a contribution to the free energy associated with the entropy of mixing of the different atomic species on each sublattice.  The configurational entropy $S_i$ associated with occupancy of sublattice $i$ is
\begin{equation*}
	S_i = k_\text{B} \ln \frac{\left(\sum_j n_{i}^{(j)}\right)!}{\prod_j \left(n_{i}^{(j)}\,!\right)},
\end{equation*}
\noindent where $n_{i}^{(j)}$ is the number of atoms of species $j$ on sublattice $i$. We can rewrite $n_{i}^{(j)}$ in terms of the concentrations $c_{i}^{(j)}$ as $n_{i}^{(j)}=c_{i}^{(j)}N_i/C_i$, where $N_i$ is the total number of atoms of type $i$. We can then rewrite this as the entropy per unit volume ($\Omega_i$ is the atomic volume on sublattice $i$) and expand it using the Stirling approximation, as
\begin{equation}
	s_i	\approx -\frac{k_\text{B}}{\Omega_i}\sum_j \frac{c_{i}^{(j)}}{C_i}  \,\ln \frac{c_{i}^{(j)}}{C_i}.
	\label{entropy_approx}
\end{equation}

The total free energy of the system is, therefore,
\begin{eqnarray}
\fl	G \left(\{\psi\}, \{c\}\right)
= \int_V &\sum_{k} \kappa_{k} \sum_{m} \left\|\mathbf{\nabla} \psi_k^{(m)}\right\|^2+
\sum_{i} \sum_{j} \lambda_{i}^{(j)} \left\|\mathbf{\nabla} c_i^{(j)}\right\|^2 +\nonumber\\
	&\sum_{k} D_k \sum_{m} \left(\psi_k^{(m)}\right)^2 \left(1-\psi_k^{(m)}\right)^2 +\nonumber\\
	&\sum_i \left[ \sum_k A_i(k) \sum_{m} \psi_k^{(m)} \right] \left(\sum_j c_i^{(j)} - C_i\right)^2+\nonumber\\
	&\sum_{k} \sum_{k^\prime < k} X_{kk^\prime} \left(\sum_m \psi_k^{(m)}\right)^2\left(\sum_{m^\prime} \psi_{k^\prime}^{(m^\prime)}\right)^2 +\nonumber\\
	&\sum_{k} X_{kk} \left[\sum_m \sum_{m^\prime<m} \left(\psi_k^{(m)}\right)^2\left(\psi_{k}^{(m^\prime)}\right)^2\right]+\nonumber\\
	&k_\text{B}\,T \sum_i \frac{1}{\Omega_i}\sum_j \frac{c_{i}^{(j)}}{C_i} \,\ln \frac{c_{i}^{(j)}}{C_i} \; dV.
\label{G_equation}
\end{eqnarray}
\noindent An elastic energy term can also be included in the free energy in order to account for possible elastic misfit between the various phases as shown in \cite{Khachaturyan-83,Hu-elasticity}.

Interfacial segregation is a common feature in most multiphase and/or polycrystalline materials.  Segregation of an element to the interface occurs if the free energy of the system is lowered by moving that element from the interior of a grain or single phase region to the interface.  In order to incorporate such an effect, we modify one of the contributions to the free energy expression (\ref{G_equation}) that controls interface energies. The appropriate enthalpy contribution is $h_3$ (interfaces) or $h_4$ (grain boundaries).  This is accomplished by replacing $X_{k k^\prime}$ with $X_{k k^\prime} - Y_{k k^\prime}(i,j)\;c_i^{(j)}$.   The set of parameters $Y_{k k^\prime}(i,j)$ describes the lowering of the $k\,k^\prime$ interface (or the grain boundaries of phase $k$ if $k=k'$) energy upon segregation of species $j$ (on sublattice $i$).

\section{Analytical results}
Before presenting a series of phase-field simulation results for the evolution of the polycrystalline silicide thin film, we examine how the general model described above describes segregation in order to confirm that the model is capable of reproducing the correct physics.  In particular, we examine the case of a grain boundary between two semi-infinite grains, $\psi^{(1)}$ and $\psi^{(2)}$ in one dimension. Since only two grains of the same phase are considered the only unknowns are the phase fraction of grain 1, $\psi(x)$, and the composition profile across the grain boundary. We assume that there is only one species per sublattice except on one sublattice where there are two elements, one of which segregates to the grain boundary. As an example, we consider the case of Ni(Pt)Si. Hence, all compositions are fixed in terms of the Pt composition profile, $c(x)$. The total energy is
\begin{eqnarray}
  G \left(\psi, c\right) = \int_V& (2D + X_\text{GB} - Y_\text{GB}\, c) \psi^2(1-\psi)^2 + (\lambda_\text{Ni}+\lambda_\text{Pt}) c'^2 +\nonumber\\
  & 2\kappa\, \psi'^2 + k_\text{B} T \left[(1-2c) \ln (1-2c) + 2c \ln 2c\right] \,\rmd V,
\label{G-at_gb}
\end{eqnarray}
\noindent where the primes indicate derivatives with respect to the space coordinate $x$.

If the segregation parameter $Y_\text{GB}$ is zero, the free energy can be split into two integrals, one depends on $\psi$ only and the other on $c$ only. Therefore, equilibrium can be found by minimizing the contributions from $\psi$ and $c$ independently. 
The solution to $\delta G/\delta \psi = 0$ is \cite{Karma-encyclopedia-01}
\begin{equation}
	\psi(x) = \frac{1}{2} \left(1+\tanh\frac{x}{W}\right),
	\label{solution_eta1_Y0}
\end{equation}
\noindent where $W$ is the width of the grain boundary:
\begin{equation}
	W^2=\frac{8\kappa}{2 D + X_\text{GB}}.
	\label{def_W}
\end{equation}
\noindent This is the canonical profile of an interface where there is no interfacial segregation. Since $Y_\text{GB}=0$, the concentration profile $c(x)$ is uniform.

\begin{figure}
\centering
    \includegraphics[width=7.5cm]{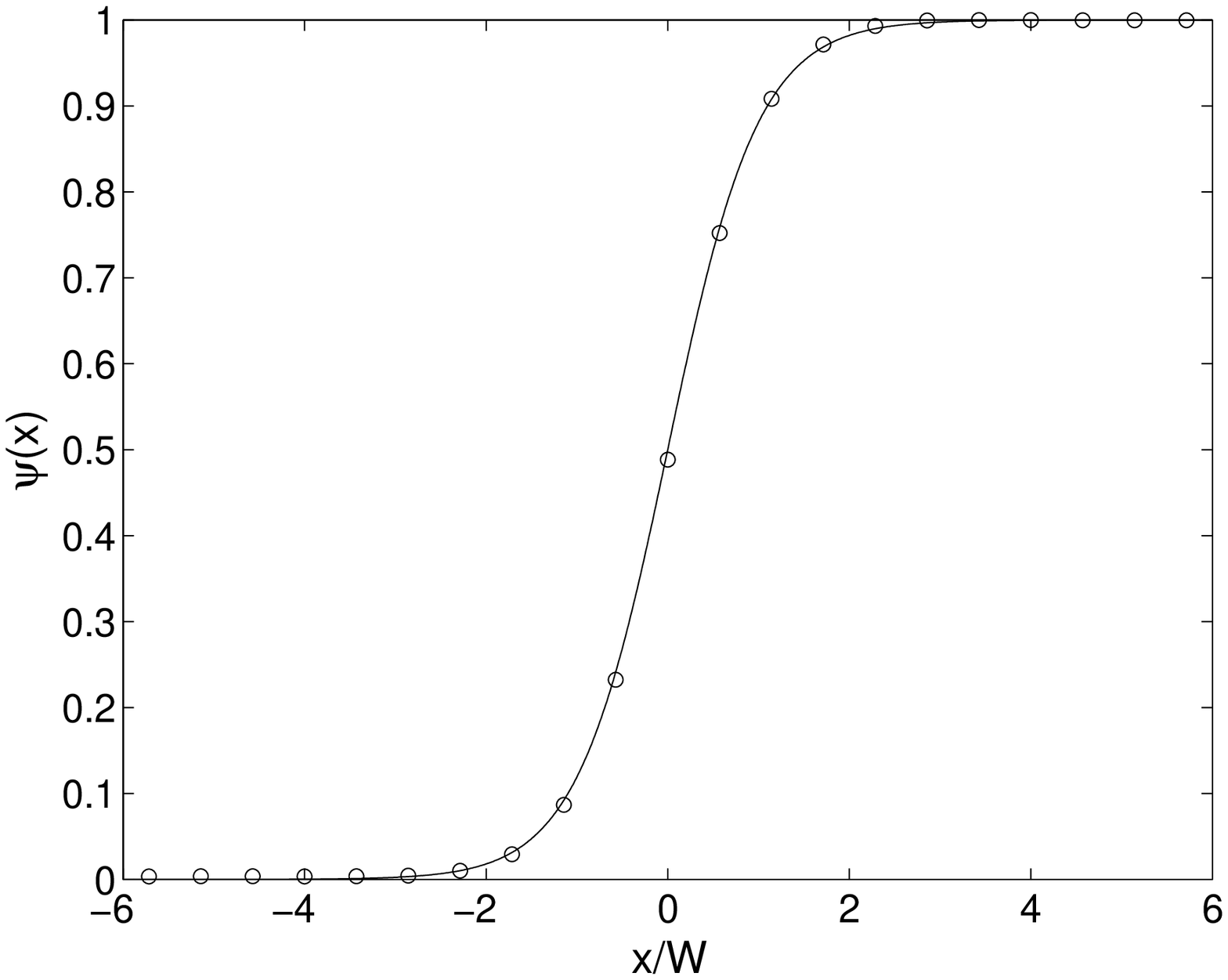}
    \includegraphics[width=7.5cm]{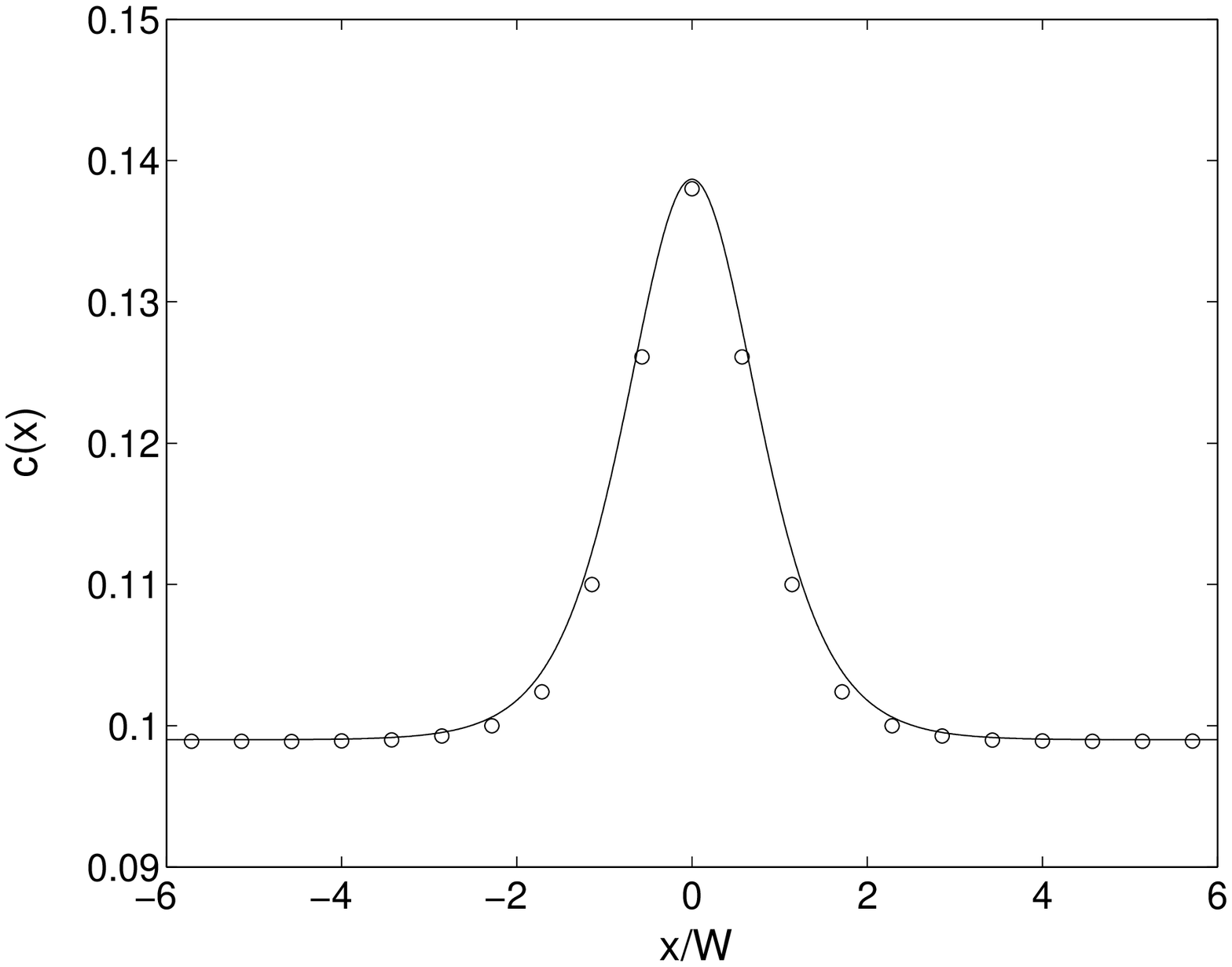}
\caption{\label{plot_epsilon} $\psi$ (left) and $c$ (right) as a function of $x/W$. The lines are obtained from (\ref{solution_eta1_Y0}) and (\ref{gamma_2}) and the circles numerically.}
\end{figure}

If $Y_\text{GB}\ne0$, the concentration profile becomes
\begin{equation}
	c(x) \approx c_0+ \frac{1}{24}\frac{Y_\text{GB}}{2D+X_\text{GB}-Y_\text{GB}\,c_0}\frac{\kappa}{\lambda_\text{Ni}+\lambda_\text{Pt}} \cosh^{-2}\frac{x}{W}
	\label{gamma_2}
\end{equation}
\noindent where $c_0$ is the nominal concentration. The derivation (and assumptions) is presented in the appendix. Figure \ref{plot_epsilon} shows $\psi(x)$ and $c(x)$ obtained from (\ref{solution_eta1_Y0}) and (\ref{gamma_2}) and from a direct numerical solution of the coupled set of equations. The results indicate that grain boundary segregation does occur, its width is determined by the grain boundary width $W$. Figure \ref{plot_epsilon} shows very good agreement between (\ref{gamma_2}) and simulations.

\section{\label{sec_results} Phase-field simulation results}

\subsection{Numerical method}
In this section we show simulation results for two-dimensional systems with periodic boundaries along both directions.
The evolution equations, (\ref{Cahn-Hillard}) and (\ref{Allen-Cahn}), are solved in reciprocal space using a semi-implicit method \cite{Chen-98}.

The direct implementation of the phase-field method is complicated by the fact that the compositions and order parameters are not {\it a priori} to lie within the physically meaningful range, i.e.\ between zero and unity. To address this issue, we introduce a smooth penalty function to the free energy when the order parameters or compositions are outside of this range:
\begin{equation}
G_\text{penalty} =	
\cases{
	\mu f^2 - \nu f		& \text{if $f < 0$,}\\
	0 				& \text{if $f \in [0,1]$,}\\
	\mu (1-f)^2 - \nu (1-f) & \text{if $f > 1$.}
}
\label{def_fence}
\end{equation}
\noindent Here $f$ can be either an order parameter or a composition and $\mu$ and $\nu$ are positive constants that are set sufficiently large to avoid non-physical values of $f$.

A similar situation exists with regard to the entropy. Expression (\ref{entropy_approx}) was derived for $c_{i}^{(j)}/C_i \in [0,\,1]$, it is undefined if $c_{i}^{(j)} \le 0$. If $c_{i}^{(j)}$ is larger than $C_i$, (\ref{entropy_approx}) is defined (although not physical) and it can therefore be used as is. We extend the entropy into the unphysical parameter space as follows:
\begin{equation}
 \frac{\partial s_i^\text{num}}{\partial c} := 
\cases{
	-k_\text{B} / (\Omega_i C_i) \left[c_i^{(j)}/(C_i\,\varepsilon) + \ln \varepsilon \right] & \text{if $c_i^{(j)} \le C_i\,\varepsilon$,}\\
	-k_\text{B} / (\Omega_i C_i) \left[ 1 + \ln \left(c_i^{(j)}/C_i\right) \right] & \text{if $c_i^{(j)} > C_i \,\varepsilon$.}
}
\label{def_entropy_derivative}
\end{equation}
\noindent Here $\varepsilon$ is a small parameter that defines the range over which the entropy is modified. Expression (\ref{def_entropy_derivative}) ensures that ${\partial s_i^\text{num}}/{\partial c}$ and its first derivative are continuous at ${c_i^{(j)} = C_i\,\varepsilon}$.

\subsection{\label{subsec-angles}Equilibrium grooving angles}
In equilibrium, a grain boundary intersects an interface at an angle $\theta$ that is determined by their relative energies (i.e., the Young-Dupree angle) \cite{Mullins}:
\begin{equation}
	\cos \theta_\text{I} = \frac{\gamma_\text{GB}}{2 \gamma_\text{I}},
	\label{def_theta}
\end{equation}
\noindent where the geometry is defined in figure~\ref{geometry}. $\gamma_\text{GB}$ and $\gamma_\text{I}$ are the grain boundary and interface energies, respectively.  This equilibrium angle should also be maintained in the immediate vicinity of the intersection even far from equilibrium, since the quantity of matter that must be transported to establish the appropriate angle is infinitesimal.  Hence, this condition is often viewed as a boundary condition.  No such condition is applied here, the angle is an outcome of the simulations.  In this section, we investigate whether the phase-field model yields the correct angular condition.

\begin{figure}
\centering
    \includegraphics{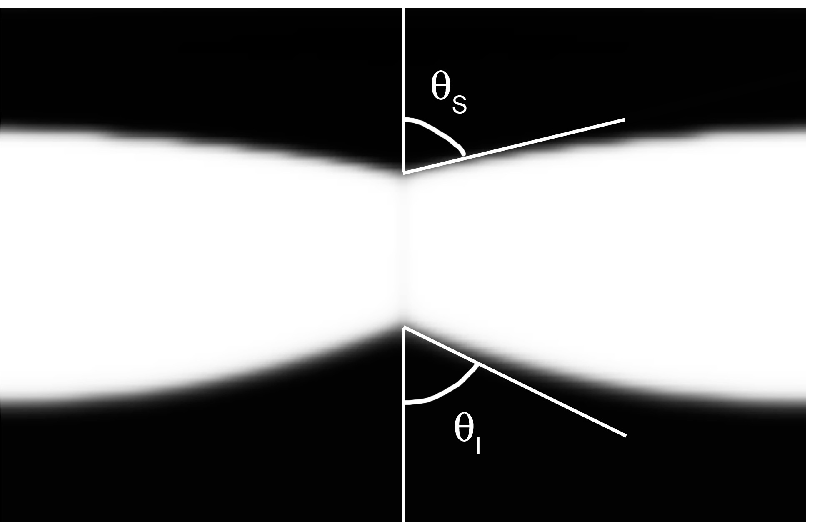}
	\hspace{.5cm}
    \includegraphics{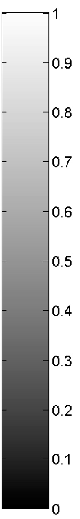}
\caption{\label{angles} A late-time image of two NiSi grains on a Si substrate (the grey scale indicates the fraction the NiSi phase present). The angles shown were obtained as $\arccos [\gamma_\text{GB}/(2\gamma_\text{I})]$ and $\arccos [ \gamma_\text{GB}/(2\gamma_\text{S})]$.}
\end{figure}

Figure~\ref{angles} shows the example of a NiSi film on a Si substrate and with a free surface. The ratio of the grain boundary to twice the interface energies is $\gamma_\text{GB}/(2\gamma_\text{I}) = 0.441$, corresponding to $\theta_\text{I}=63.9^\circ$  and the ratio of the grain boundary to twice the surface energy is $\gamma_\text{GB}/(2\gamma_\text{S}) = 0.235$, corresponding to $\theta_\text{S}=76.4^\circ$. The angles drawn in figure~\ref{angles} correspond to these theoretical angles.  Clearly, the phase-field simulations accurately reproduce the theoretical predictions of (\ref{def_theta}). 

\subsection{\label{subsec-Pt_segreg}Platinum segregation}
\begin{figure}
\centering
\subfigure 
{\label{segreg_GB(a)}
	\includegraphics[width=7cm]{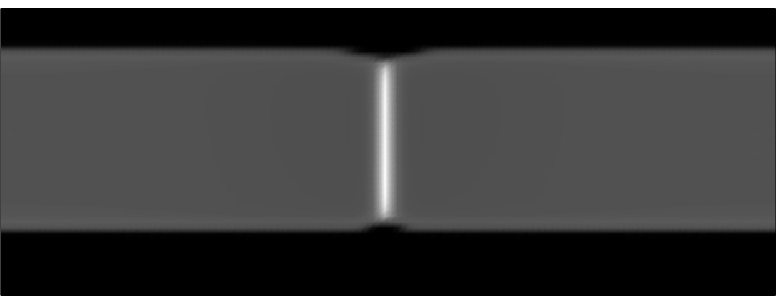}}
{\label{segreg_GB(b)}
	\includegraphics[width=7cm]{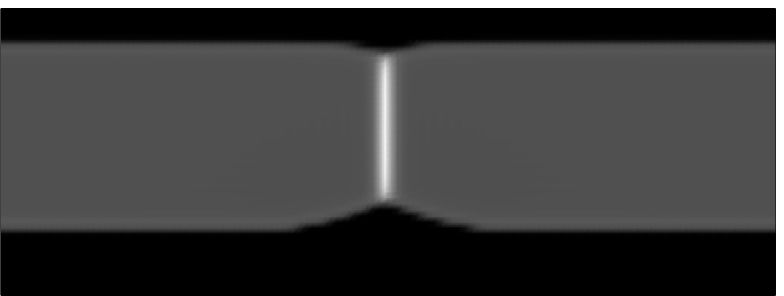}
	\includegraphics[width=.85cm]{buf.eps}}
{\label{segreg_GB(c)}
	\includegraphics[width=7cm]{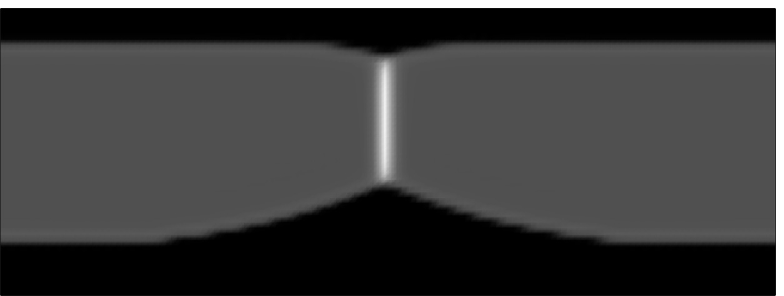}}
{\label{segreg_GB(d)}
	\includegraphics[width=7cm]{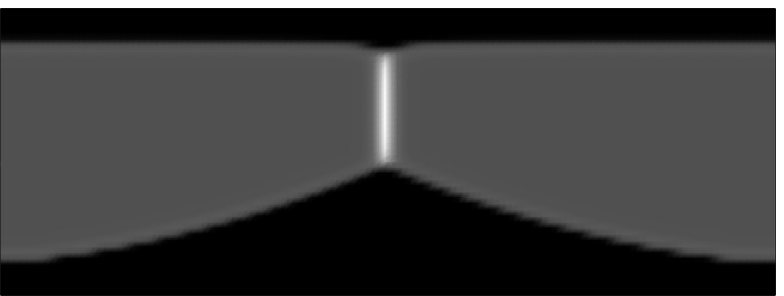}
    \includegraphics{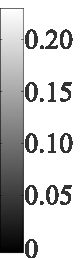}}
\caption{\label{segreg_GB} The Pt composition field at $t = 10$, $100$, $500$, and $2000$ for the case of $Y_\text{GB}=43$ (i.e., $X_\text{GB}=12-43\,c_\text{Pt}$).  The other parameters used for both simulations were $\lambda_\text{Si} = 10$, $\lambda_\text{Ni} = 3$, $\lambda_\text{Pt} = 3$, $\kappa_\text{NiSi}=2$, $\kappa_\text{vacuum}=5$, $k_\text{B} T/\Omega= 0.4$, $X_\text{Si-NiSi}=16$, $X_\text{NiSi-vacuum}=20$ and $A = 20$ for all phases.}
\end{figure}

Figure~\ref{segreg_GB} shows the Pt concentration profiles in two Ni(Pt)Si grains of a thin film on a Si substrate at different times. As $Y_\text{GB}\ne 0$ segregation occurs.
The platinum composition at the interface, three times as large as the nominal composition, is established very early in the simulations. A similar approach can be employed to model Pt segregation to other interfaces, e.g., choosing a non-zero value for $Y_\text{Si-NiSi}$ and $Y_\text{vacuum-NiSi}$ (i.e., Pt concentration-dependent $X_\text{Si-NiSi}$ and $X_\text{vacuum-NiSi}$).

Figure \ref{segreg_GB} shows that the angles are established early in the simulations and are maintained even when the surfaces and interfaces do not have constant curvatures (i.e., away from equilibrium) as proposed in section \ref{subsec-angles}.

\section{Conclusion}
We presented a phase-field model for grain boundary grooving in polycrystalline thin films on a substrate. The model is quite general, describing compounds and solid solution alloys with sufficient freedom to choose solubilities, grain boundary and interface energies, and heats of segregation to all interfaces.   
We also derived analytical expressions for describing interface profiles ---with and without segregation--- and we confirmed these using numerical simulation. The present model was shown to accurately reproduce the theoretical grain boundary groove angles both at and far from equilibrium.   As an example, we applied the phase-field model to the special case of a Ni(Pt)Si thin film on an initially flat Si-substrate.  The present model is suitable for describing grain boundary grooving and thin film agglomeration in real systems.

\ack
The authors gratefully acknowledge the support of the A*Star Visiting Investigator Program.

\appendix
\section*{\label{appendix}Appendix: grain boundary profile}
\setcounter{section}{1}
In this appendix, we present the derivation of the composition profile at a grain boundary.
We consider the case of a grain boundary between two semi-infinite grains, $\psi^{(1)}$ and  $\psi^{(2)}$ in one dimension. Since $\psi^{(1)}+\psi^{(2)}=1$ (only two grains of the same phase are considered) the only unknowns are the phase fraction of grain 1, $\psi(x):=\psi^{(1)}(x)=1-\psi^{(2)}(x)$, and the compositions. We assume that there are two species on each sublattice $i$ and that $c_{i}^{(1)} + c_{i}^{(2)} = C_i$ for all $i$. Hence, all compositions are fixed in terms of one composition profile per sublattice, $c_i(x)$. The total energy is
\begin{eqnarray}
\fl  G \left(\psi, \left\{c_i\right\}\right) = \int_V& \left(2D + X_\text{GB} - \sum_i Y_i\, c_i\right) \psi^2(1-\psi)^2 + \sum_i\left(\lambda_i^{(1)}+\lambda_i^{(2)}\right) \left(c_i'\right)^2 +\nonumber\\
\fl  & 2\kappa\, (\psi')^2 + k_\text{B} T \sum_i \frac{1}{\Omega_i} \left[\frac{c_i}{C_i} \ln \frac{c_i}{C_i} + \left(1-\frac{c_i}{C_i}\right) \ln \left(1-\frac{c_i}{C_i}\right) \right] \,\rmd V,
\label{G-at_gb-A}
\end{eqnarray}
\noindent where $Y_i$ controls the segregation to the grain boundary for atoms of type $i$.

The functional derivatives of the free energy with respect to $\psi$ and $c_i$ are
\begin{eqnarray}
	\frac{\delta G}{\delta \psi} = 2\left(2 D + X_\text{GB} - \sum_i Y_i\, c_i \right) \psi (1-\psi)(1-2\psi) -4\,\kappa \,\psi''\label{derivatives-G-psi-A}\\
	\frac{\delta G}{\delta c_i} = \frac{k_B T}{C_i\,\Omega_i} \,\ln \frac{c_i}{C_i-c_i} - Y_i \,\psi^2 (1-\psi)^2 - 2 \left(\lambda_i^{(1)}+\lambda_i^{(2)}\right)c_i''.
\label{derivatives-G-c-A}
\end{eqnarray}
\noindent The only dependence of ${\delta G}/{\delta \psi}$ on the compositions is from $\sum_i Y_i\, c_i$. If the deviation of $c_i$ from the far field composition is small, then $\sum_i Y_i\, c_i(x) \approx \sum_i Y_i\, c_i(\infty)$ and 
\begin{equation}
	\psi(x) \approx \frac{1}{2} \left(1+\tanh\frac{x}{W}\right),
	\label{solution_eta1_Y0-A}
\end{equation}
\noindent with
\begin{equation}
	W^2 = \frac{8\kappa}{2 D + X_\text{GB} - \sum_i Y_i\, c_i(\infty)}.
	\label{def_W-A}
\end{equation}
\noindent If there is no segregation, all $Y_i$ are zero and we recover (\ref{solution_eta1_Y0})  and (\ref{def_W}). 

The equilibrium on sublattice $i$, given by ${\delta G}/{\delta c_i}=0$, is independent of the other sublattices. We can therefore limit ourselves to one sublattice without loss of generality. We will call $c$ the composition of species 1 on this sublattice. If the deviation of this composition profile from the nominal composition $c_0$ is small, we can describe the profile as $c(x) = c_0 +\gamma(x)$ with $|\gamma(x)| << c_0$. 
Uding the expression from (\ref{solution_eta1_Y0-A}) for $\psi$ in (\ref{derivatives-G-c-A}), to first order in $\gamma$, $\delta G/\delta c = 0$ gives
\begin{equation}
	2\left(\lambda^{(1)}+\lambda^{(2)}\right) \gamma''(x) - \frac{k_\text{B}\, T}{\Omega}\frac{1}{c_0(C-c_0)} \gamma(x) \approx -\frac{Y}{16} \cosh^{-4}\frac{x}{W}.
\end{equation}
\noindent This equation is of the form $W^2 \gamma''(x) - \alpha^2 \gamma(x) = - \delta \cosh^{-4}(x/W)$, where $\alpha$ and $\delta$ are dimensionless constants. The solution to this equation is
\begin{eqnarray}
	c(x) \approx &c_0+\frac{\delta}{6} \cosh^{-2}\frac{x}{W} + A B\exp\frac{\alpha x}{W} +A (1-B) \exp\frac{-\alpha x}{W}+ \nonumber \\
		&\frac{2}{3} \delta \sum_{k=1}^{+\infty} {(-1)^k k \frac{\alpha^2-4}{\alpha^2-4k^2}}\left[B\exp\frac{2k x}{W} +(1-B) \exp\frac{-2k x}{W}\right]
	\label{gamma_gal}
\end{eqnarray}
\noindent where $A$ and $B$ are integration constants.

Since the general solution given in (\ref{gamma_gal}) is complicated we rewrite it for two particular values of  $\alpha$. For $\alpha = 1$ 
\begin{equation}
\fl	c(x) \approx c_0+\frac{\delta}{2} \left[\frac{1}{3} \cosh^{-2}\frac{x}{W} +\frac{\pi}{2} \cosh{\frac{x}{W}} - 1 - \left(\sinh \frac{x}{W}\right) \arctan \left(\sinh{\frac{x}{W}}\right)\right],
	\label{gamma_1}
\end{equation}
\noindent and for $\alpha = 2$
\begin{equation}
	c(x) \approx c_0+\frac{\delta}{6} \cosh^{-2}\frac{x}{W}.
	\label{gamma_2A}
\end{equation}
\noindent Equation~(\ref{gamma_2}) is written for $\alpha=2$. Figure \ref{plot_epsilon} shows $\psi(x)$ from (\ref{solution_eta1_Y0}) and $c(x)$ from (\ref{gamma_2}), i.e.\ (\ref{gamma_2A}), as functions of $x/W$. 
The circles correspond to simulations using a set of parameters such that $\alpha=2$.

\end{document}